# REFERENCE ARCHITECTURE FOR SMAC SOLUTIONS

Shankar Kambhampaty[1] and Sasirekha Kambhampaty[2]

[1]Computer Science Corporation (CSC), India
skambhampaty@gmail.com

[2]Student, Department of Computer Science, GRIET, Hyderabad, India
ksasirekha@gmail.com

## ABSTRACT

*Web and internet computing is evolving into a combination of social media, mobile, analytics and cloud (SMAC) solutions. There is a need for an integrated approach when developing solutions that address web scale requirements with technologies that enable SMAC solutions. This paper presents an architecture model for the integrated approach that can form the basis for solutions and result in reuse, integration and agility for the business and IT in an enterprise.*

## 1. INTRODUCTION

The rise of mobile apps has been causing increasing demand for accessing functionality from outside the infrastructure of enterprises. Consumers are demanding information relevant to their context anywhere, anytime, on any device. At the same time, the rise of "Internet of Things" also requires a standard interface for devices to communicate the data captured by different types of devices (such as refrigerators & dish washers) through published interfaces. Additionally, social media applications (Facebook, Twitter etc.) are fast becoming access channels for consumers to interact with products (e.g. Salesforce CRM) and services (such as internet banking) of enterprises. All these taken together constitute web-scale demand that needs to be addressed by solutions of enterprise.

Web and internet computing technologies are making significant advances with evolving technologies that may be grouped under four categories – social media, mobile, analytics and cloud computing [1]. Enterprises employ these technologies to deal with the web-scale demand when offering products and services to their customers.

The problem, however, is that point solutions are being provided in several enterprises without taking a holistic view of the current and future needs of the enterprise leading to "patch work" of the IT in the enterprise thereby increasing the cost in the long run.

The purpose of this paper is to provide reference architecture to address the above problem and enable in reuse, integration and agility thereby reducing the total cost of ownership for the enterprise.

## 2. WEB-SCALE DEMAND

"Web-scale describes the tendency of modern architectures to grow at (far-) greater-than-linear rates. Systems that claim to be Web-scale are able to handle rapid growth efficiently and not have bottlenecks that require re-architecting at critical moments" [2].

Some of the sources of web-scale demand are as follows:

1) Data from sensors (such as electricity meters) to applications that manage business processes (e.g. power load management).

2) Consumer devices (such as wearables) that provide data on continuous basis to remote healthcare monitoring systems.

3) Data feeds from social media sites that provide insights on consumer buying behaviour.

4) Mobile devices accessing news sites when sensational events happen round the world.

Engineering firms that deal with large number of sensor data and enterprises that provide products and services to a substantial volume of consumers are likely to experience the web-scale demand. Gartner predicts that "By 2017 Web-Scale IT Will Be an Architectural Approach Found Operating in 50 Percent of Global Enterprises" [3]. This is because of the disruption in how, traditionally, business has been conducted and the trend towards digital business models.

## 3. SMAC SOLUTIONS

Several key technology advancements have been taking place in social media, mobile, analytics and cloud computing that are collectively referred to by the term SMAC.

Social media platforms are being leveraged by enterprises as one of the access channels for customers. For instance retail banks (such as ICICIBank) provide apps on social media platforms (e.g. Facebook) to allow customers to perform the day-to-day banking transactions. This requires business solutions of the enterprise (e.g. core banking solution) to be integrated with the social media platform to enable large number of transactions to flow through the social media channel [4]. Social media platforms being open to all, can potentially result in millions of transactions that have to be handled in a secure manner over relatively short periods of time.

Mobile platforms are presenting both a huge opportunity and significant challenge to large enterprises. It is necessary to provide mobile apps to enterprise users and customers to interact with the enterprise for products and services while supporting a wide variety of mobile operating systems (iOS, Android, Windows etc.) that run on a multitude of devices with varying form factors. The mobile apps have to be continuously kept up-to-date and have to interact with the core business solutions of the enterprise in a secure and a user-friendly manner. The back-end solutions for the mobile enables need to be highly scalable, insulated against the regularly changing front-end mobile apps. An MBaaS platform, mobile backend as a service, may also be part of solution mix to support integration and address the mobile notification requirements of the enterprise [5].

Analytics solutions typically extract intelligence from structured and unstructured data to help enterprises make business decisions. The business enterprises require real-time decisions to be made based on the large amount of business and transaction data that flows from the social media and mobile applications. Big data solutions based on technologies such as Storm and Hadoop are central to organization IT landscape for this purpose [6].

Cloud platforms and related solutions provide the back-end infrastructure to host and deploy applications that support business in addition to traditional IT within enterprises. The cloud platforms are enabling scalability and reducing total cost of ownership (TCO) through pay-per-use cost model while providing one or more of the following "as a service" offerings [7]:
- Infrastructure as a Service (IaaS)
- Platform as Service (PaaS)
- Software as a Service (SaaS)

## 4. NEED FOR INTEGRATED APPROACH

In most enterprises, the solutions for each of the SMAC technologies are being led and managed by different teams from various units without taking into account an enterprise-level view. Often, this is because the driving factors for the solutions for each of the SMAC technologies are different and justification for business investments and return on investment (ROI) are measured differently. While an enterprise architecture view may exist in many enterprises and SMAC technologies may be formally be identified as strategic in their roadmaps, the fact that they are managed by different groups in the enterprise with goals that are not necessarily aligned does not foster an integrated model for the SMAC solutions. Consequently, SMAC solutions in enterprises have evolved to be disparate systems with little or no integration resulting in efficiencies, increased total cost of ownership (TCO) and lesser time to market [8].

An integrated approach with a common layer acting as "broker" for social media, mobile, analytics and cloud platforms can enable scalable (due to cloud support), customer facing (due to social media and mobile touch points) and intelligent (drawing from analytics system interfaces) solutions. An example of such a solution is a travel solution that gathers intelligence related to customer buying behaviour from social media and airline web sites and pushes alternate or related product purchasing options to clients through mobile apps. Another example is a power company that switches users to different power providers based on load, time of the day and pricing using data from sensors in the power meters and informing users through mobile text messages.

## 5. SERVICES MODEL – THE FOUNDATION

The services model that provides loose coupling of service consumers with service providers and integrated with a "broker" pattern, provides a good foundation for an integrated architecture model for SMAC solutions [9]. Equally important is the fact that most enterprises with mature IT practices have made substantial investments in implementations of Service-Oriented Architecture (SOA).

A generic services model for an enterprise may be defined based on four types of services, namely, *activity services* (A), *business process services* (B), *client services* (C) and *data services* (D) [10]. Figure 1 depicts such an enterprise-level generic architecture based on the four types of services.

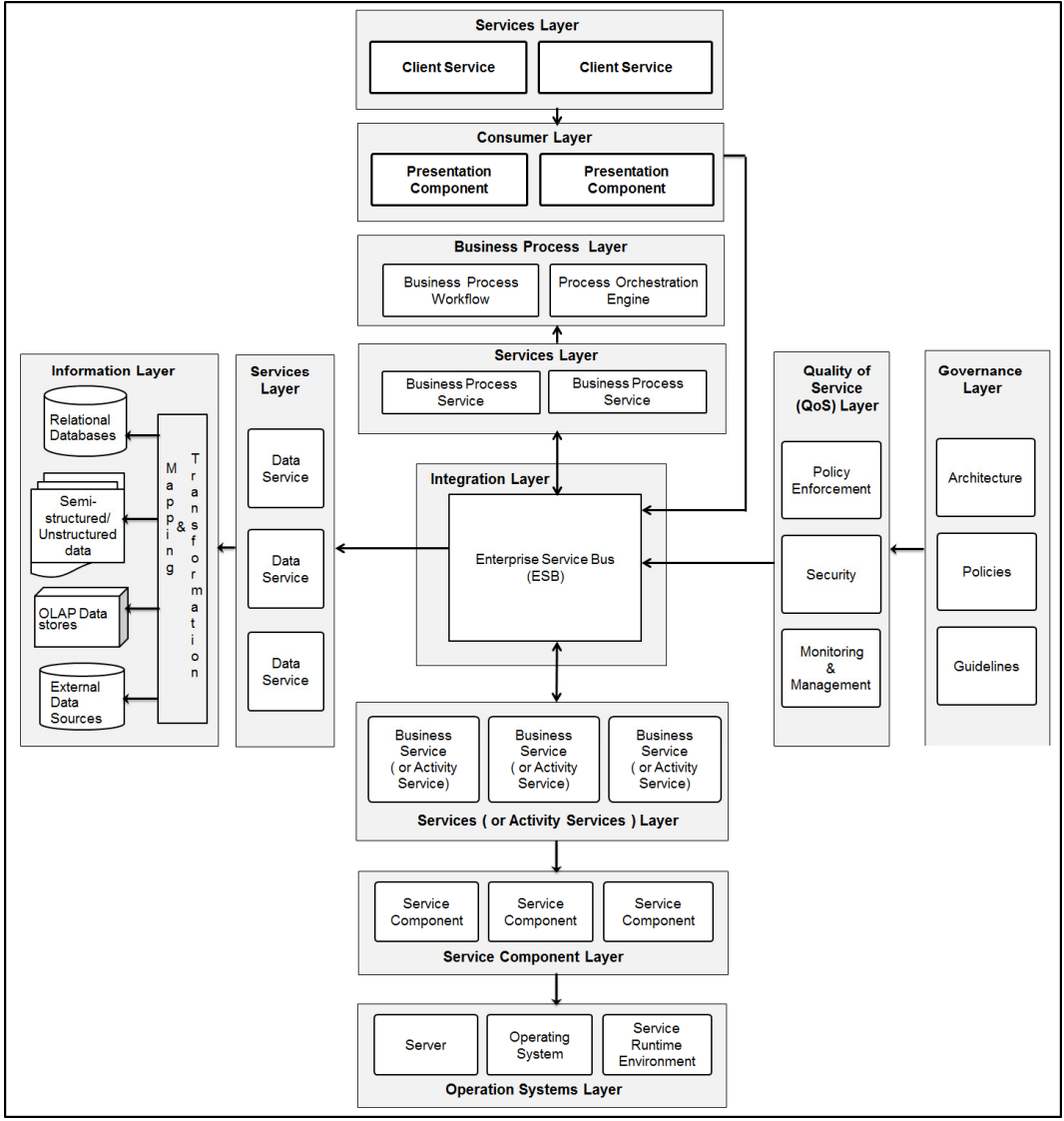

Figure 1: Enterprise-level generic architecture based on services model

The generic architecture shown in Figure 1 represents a logical perspective of the four types of services and their relationships.

A description of the services is as follows:

1. **Activity services (or Business services)**: The activity services encapsulate functionality exposed as services in various business applications. These services are reusable business level services that can be orchestrated as part of a configured business process. These services will be referred to as type "A" services.

2. **Business process services:** Business process services handle workflow requirements (e.g. approval processes) through orchestration of business processes, each step of which is implemented as a service. These services enable externalization of business processes through their specification as workflows and orchestration by process orchestration engine resulting in agility for the enterprise. They will be referred to as type "B" services.

3. **Client services:** Client services provide content meant for "front-end" applications of the enterprise to enable clients/partners to access their business data (e.g. order information) and enterprise users to get an aggregated view (e.g. enterprise dashboard). APIs defined by an enterprise for consumption by clients/partners may be implemented with this category of services. These services will be referred to as type "C" services.

4. **Data services**: Data services provide access to data in various sources. There are typically two types of data services that have to be provided in an enterprise:
- Structured data stores (e.g. relational databases)
- Unstructured data stores (e.g. Big data data stores that run analytics)

These services will be referred to as type "D" services.

The Enterprise Service Bus (ESB) in the integration layer enables a smooth communication between the service providers and service consumers. An Enterprise Service Bus (ESB) pattern abstracts the mediation and interaction elements needed for communication on a bus that is used for integration of services.

The generic architecture in Figure 1 also addresses the need in enterprises to define policies for key non-functional requirements and their enforcement through elements in governance and quality of service (QoS) layers:

1. Governance – Service interaction (service calls from service consumers to service providers) has to be governed. To make this happen, policies are defined in governance layer for non-functional requirements such as security taking into account key performance indicators for business and SLAs for IT.

2. Quality of Service (QoS) – Service interaction has to be monitored and secure and in accordance with the policies defined. This is ensured through monitoring of service interactions (application monitoring, business activity monitoring and IT systems monitoring) and enforcing of policies at runtime.

## 7. REFERENCE ARCHITECTURE FOR SMAC SOLUTIONS

The emergence of APIs as human readable, externally facing, light-weight services has made APIs central to solutions based on SMAC technologies. In fact, the one aspect common to social media, mobile, analytics and cloud solutions is access of functionality through APIs exposed by the respective platforms. Social media sites (such as Twitter) provide APIs for enabling solutions that respond to user posts (or tweets). Mobile apps consume content by making calls to APIs exposed by back-end applications in the enterprise. Big Data solutions are headed towards providing "intelligence as a service" for use by applications across the enterprise. Thus, APIs can serve to be the "glue" for the integrated architecture model.

With the APIs gaining rapid adoption, technologies that support API solutions such as API Gateways have also come into existence that enable integration and play the role of a "broker" for solutions based on SMAC technologies.

Building on the services model based generic architecture, described in the earlier section, and applying the considerations of web scale demand and trends in SMAC technologies, Reference Architecture may be defined for SMAC solutions as shown in Figure 2.

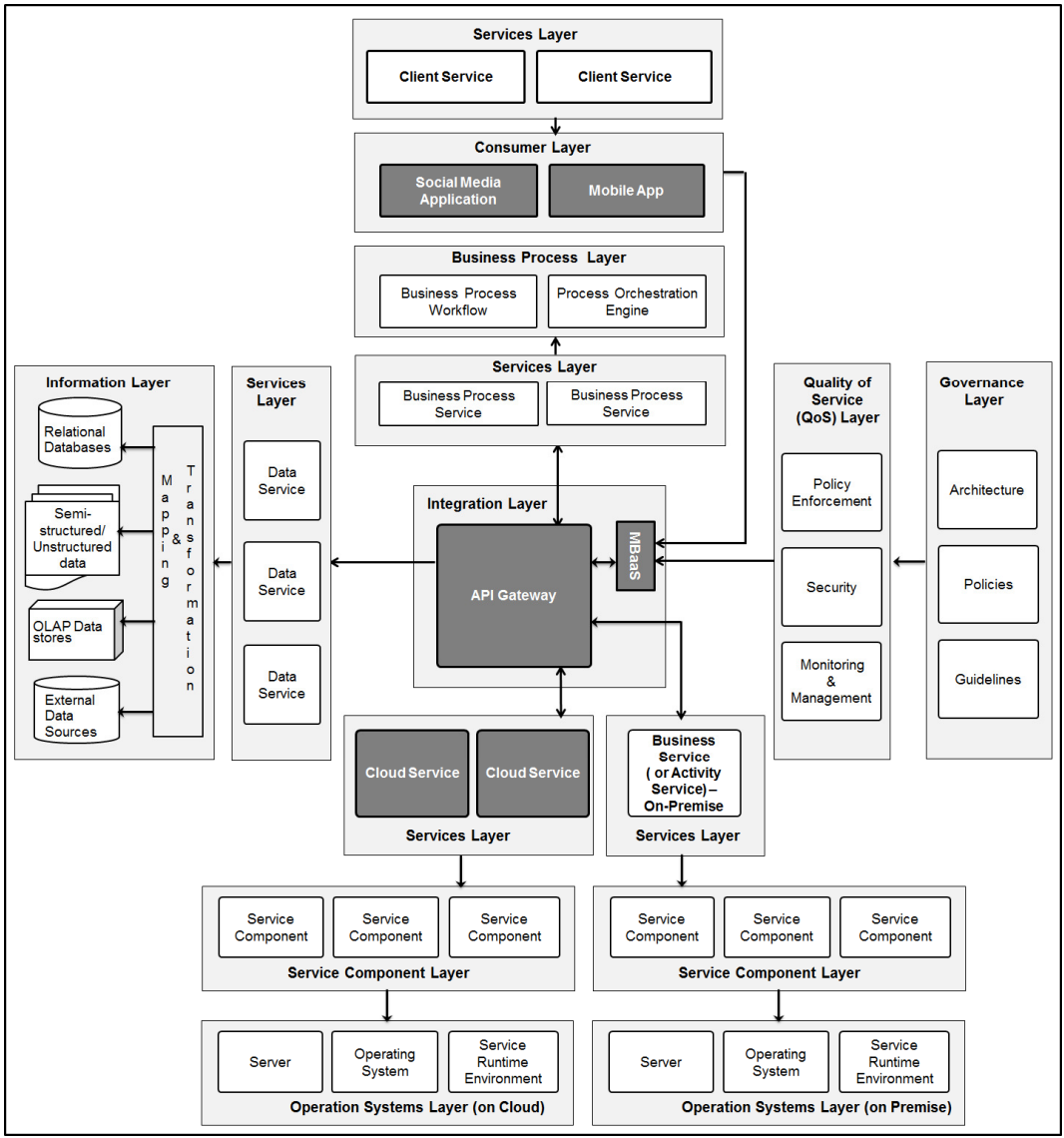

Figure 2: Reference Architecture for SMAC solutions

The Reference Architecture for SMAC solutions depicted in Figure 2 brings together the social media, mobile, analytics and cloud solution components through exposing their functionality through services, integrating them through integration layer and orchestrating them through the business process layer.

The consumer layer in Figure 2 shows social media client applications and mobile apps making API calls to the integration layer. *API Gateway* and *Mobile Backend as a Service* (MBaaS) platform are key elements of the integration layer. API Gateway is a "broker" pattern that exposes a standard set of APIs defined to support both intranet applications as well social media and mobile clients. SMAC solutions expose light-weight APIs (which are RESTful web services) instead of SOAP based web services and thus the API Gateway can effectively take the place of an ESB. Enterprises that have significant number of SOAP based web services (or non-customer facing applications) and have made substantial investments in ESB infrastructure may continue to use the ESB for service integration. But given the trend in the industry, the API

Gateway is expected to play a key role in integration of SMAC solutions. Another important element in the integration layer is the MBaaS platform that handles the needs of social media and mobile clients, especially by providing push notifications to them. The webscale demand generated by social media and mobile clients are handled by the combination of MBaaS and API Gateway elements in the integration layer.

Analytics applications are part of the information layer and Figure 2 shows data services being exposed to support the requirements of applications in the enterprise. Of particular relevance in this context are the Big data based analytics applications that will see widespread adoption in future. The webscale demand by social media and mobile clients may generate large volume of data relevant to analytics that may be captured by Big data applications as the API invocations are made to the integration layer. The analytics applications enable business intelligence decisions to be made and data services provide the means of querying the analytics applications through service calls in order to make business decisions.

Enterprises are increasingly expected to move towards hybrid cloud model in future. This would mean a certain amount of business functionality would be implemented in cloud applications and the rest in on-premise applications. Figure 2 shows elements to support both cloud and on-premise applications and their services.

The API Gateway forms the "binding glue" bringing all the services/APIs together to support improved agility and dynamism and reducing the total cost of ownership through reuse and integration [11].

The Reference Architecture so defined provides a consistent and integrated approach to addressing the needs of SMAC solutions.

## 8. IMPLEMENTATION OF REFERENCE ARCHITECTURE

The Reference Architecture for SMAC solutions is best suited to be implemented at enterprise-level. To that end, the architecture groups within enterprises may consider its adoption as part of their enterprise architecture initiatives.

The following steps are recommended in implementing the Reference Architecture:

1) Review of the IT Roadmap of the enterprise.

2) Determination of strategic importance of SMAC technologies in meeting the organization goals.

3) Assessment of maturity of implementation of services model within the enterprise.

4) Identification of products to implement API Gateway and MBaaS platforms.

5) Development of business case to justify the investment required for implementation.

6) Execution of pilot to provide the technology choices and ROI.

## 9. CONCLUSIONS

SMAC technologies are seeing increased adoption in web and internet computing. On account of webscale demand and changing needs of organizations, an integrated approach is needed in architecting and implementing SMAC solutions. The services model provides the right foundation for defining a Reference Architecture with an integrated approach. The Reference Architecture proposed in the paper can enable reuse, agility and integration to reduce the total cost of ownership and provide a competitive Enterprise IT architecture to an organization.

## REFERENCES

[1] Evans, N (2013) *SMAC and the evolution of IT*, Computerworld http://www.computerworld.com/article/2475696/it-transformation/smac-and-the-evolution-of-it.html

[2] Leibovici, A (2014) *Understanding Web-Scale Properties*, Nutanix, http://www.nutanix.com/2014/03/11/understanding-web-scale-properties

[3] Gartner (2014) Press Release, Gartner, http://www.gartner.com/newsroom/id/2675916

[4] ISACA (2010) Social Media: Business Benefits and Security, Governance and Assurance Perspectives, ISACA, http://www.isaca.org/groups/professional-english/security-trend/groupdocuments/social-media-wh-paper-26-may10-research.pdf

[5] Pezzini, M., Guttridge, K., Thoo, E., Thomas, A. (2015) Support Multiple Integration Patterns in Your Mobile App Architecture Strategy, Gartner Document ID: G00270938, http://www.gartner.com/resources/270900/270938/support_multiple_integration_270938.pdf

[6] Wähner, K. (2014) Real-Time Stream Processing as Game Changer in a Big Data World with Hadoop and Data Warehouse, InfoQ article, http://www.infoq.com/articles/stream-processing-hadoop

[7] Kambhampaty, S (2010), *Service Oriented Architecture for Enterprise and Cloud Applications*, 2nd Edition, Wiley India publication, ISBN: 978-81-265-1989-7, http://www.amazon.in/Service-oriented-Architecture-Enterprise-Cloud-Applications/dp/8126519894

[8] Overby, S (2014), How to Tame Social, Mobile, Analytics and Cloud Multisourcing, CIO, http://www.cio.com/article/2451068/outsourcing/how-to-tame-social-mobile-analytics-and-cloud-multisourcing.html

[9] Kambhampaty, S., Chandra, S. (2006), *Service oriented architecture for enterprise applications*, SEPADS'06: Proceedings of the 5th WSEAS International Conference on Software Engineering, Parallel and Distributed Systems, http://dl.acm.org/citation.cfm?id=1365739.1365748&coll=DL&dl=GUIDE&CFID=528744429&CFTOKEN=40603217

[10] Kambhampaty, S. (2007), *Service oriented analysis and design process for the enterprise*, ACS'07 Proceedings of the 7th Conference on 7th WSEAS International Conference on Applied Computer Science - Volume 7, http://dl.acm.org/citation.cfm?id=1348171.1348235&coll=DL&dl=GUIDE&CFID=528744429&CFTOKEN=40603217

[11] Oliver, A. (2014), APIs will glue together the Internet of things, Javaworld, http://www.javaworld.com/article/2606963/java-me/apis-will-glue-together-the-internet-of-things.html

**Authors**

Shankar Kambhampaty is a CSC Distinguished Architect and Chief Technology Officer (CTO) for a major account at CSC, Hyderabad, India. He is author of several international conference papers including the book titled, Service Oriented Architecture for Enterprise and Cloud Applications, Wiley India publication.

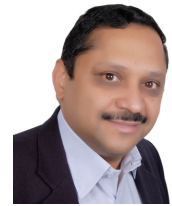

Sasirekha Kambhampaty is a student of Bachelor of Technology (B.Tech.) programme in Computer Science & Engineering at Gokaraju Rangaraju Institute of Engineering & Technology (GRIET), Hyderabad, India.

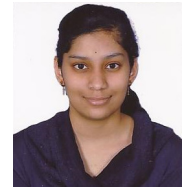